\documentclass[12pt]{iopart}

\usepackage{iopams}  

\usepackage{epsfig}

\newcommand{\ignore}[1]{}
\begin{document}

\title[Analytical Solution of the Voter Model on Disordered Networks]{Analytical Solution of the Voter Model on Disordered Networks}

\author{Federico Vazquez and V\'{i}ctor M. Egu\'{i}luz}

\address{IFISC\footnote{http://ifisc.uib.es}, Instituto de F\'{i}sica
Interdisicplinar y  Sistemas Complejos (CSIC-UIB), E-07122 Palma de Mallorca,
Spain} \ead{federico@ifisc.uib.es}

\date{\today}

\begin{abstract}
We present a mathematical description of the voter model dynamics on
heterogeneous networks.  When the  average degree of the  graph is  $\mu \leq
2$ the system reaches complete order exponentially fast.  For $\mu >2$, a
finite  system falls, before it fully orders, in  a quasistationary state in
which the average density of active links (links between opposite-state nodes)
in surviving runs is constant and equal to $\frac{(\mu-2)}{3(\mu-1)}$, while
an infinite large system stays ad infinitum in a partially ordered stationary
active state.  The mean life time of  the quasistationary state is
proportional to the mean time to reach the fully ordered state $T$, which
scales as  $T \sim \frac{(\mu-1) \mu^2 N}{(\mu-2)\,\mu_2}$,  where $N$ is the
number of nodes of the network, and $\mu_2$ is the second  moment of the
degree distribution.  We find good agreement between these  analytical results
and numerical simulations on random networks with various degree
distributions.
\end{abstract}


\maketitle

\section{Introduction}
\label{intro}

The \emph {voter model} has become one of the most popular interacting
particle systems \cite{Holley75,Liggett} with applications to the study of
diverse processes like opinion formation \cite{SanMiguel05,Castellano07},
kinetics of heterogeneous catalysis \cite{Krapivsky92,Krapivsky96}, and
species competition \cite{Clifford73}.   The general version of the model
considers a network formed by nodes holding either spin 1 or -1. In a single
event, a randomly chosen node adopts the spin of one of its neighbors, also
chosen at random.   Beyond this standard version, several variations of the
model have been considered in  the literature, to account for zealots or
inhomogeneities (individuals that favor one of the states) \cite{Mobilia03},
constrained interactions \cite{Vazquez03}, non-equivalent states
\cite{Castello07b}, asymmetric transitions  or bias \cite{Antal06}, noise
\cite{Medeiros96} and ecological diversity \cite{Durrett96}.  It is also known
that several models presenting a coarsening process without  surface tension
belong to the voter model universality class \cite{Dornic01}.

In a regular lattice, the mean magnetization, i.e., the normalized difference
in the number of $1$  and $-1$ spins, is conserved at each time step. Thus the
magnetization is not a useful order parameter to study the ordering dynamics
of the voter model.  Instead, it is common in the physics literature to use as
a order parameter the density of interfaces $\rho$, i.e, the fraction of links
connecting neighbors with opposite spins.  In a finite system, the only
possible final state is the fully ordered state,  in which all spins have the
same value, either $-1$ or $1$, and therefore all  pair of neighbors are
aligned ($\rho=0$). These are absorbing configurations  given that the system
cannot escape from them once they are reached \cite{Hammal05}.  Despite its
non-trivial dynamics, an exact solution has been obtained for regular lattices
of general dimension  $d$ \cite{Krapivsky92,Krapivsky96}, becoming one of the
few non-equilibrium models which are exactly solvable in any
dimension. Indeed, the correspondence between the voter model and a system of
coalescing random walkers  helps to solve analytically many features of the
dynamics \cite{Cox86,Scheucher88}. For $d \leq 2$, there is a coarsening
process where the average size of ordered regions composed by sites holding
the same spin continuously grows. In the thermodynamic limit, the approach to
the final frozen configuration is characterized by the monotonic decrease in
$\rho$, that decays as $\rho \sim t^{-1/2}$ in $1d$ and $\rho \sim (\ln
t)^{-1}$ in $2d$ \cite{Krapivsky92}. For $d>2$,  the density of active
interfaces behaves as $\rho(t) \sim a-b\,t^{-d/2}$ \cite{Krapivsky96}, thus
$\rho (t)$ reaches a constant value in the long time limit where the system
reaches a stationary active state with nodes continuously flipping their
spins.  That is to say, full order is never reached.  We need to clarify that
the last is only true for infinite large systems, given that fluctuations in
finite size lattices make the system to ultimately reach complete order. The
level of order in the stationary state is quantified by the two-spin
correlation function $C_{ij} \equiv \langle S_i S_j \rangle$ between spins $i$
and $j$, that decays with their spatial separation $r=|i-j|$ as $C(r) \sim
r^{(2-d)}$ \cite{Redner01}, i.e, far apart spins become uncorrelated.   Recent
studies of the voter model on fractals with fractal dimension in the range
$(1,2)$, reveal that the system orders following $\rho (t) \sim t^{-\alpha}$,
with the exponent $\alpha$ in the range $(0,1)$ \cite{Suchecki06,Bab08}.

The voter model has recently been investigated on complex networks
\cite{Castellano03,Vilone04,Suchecki05b,Sood05,Castellano05,Suchecki05a,Castello07a}, where its behavior seems to strongly depend on the topological
characteristics of the network.  A peculiar aspect is that the dynamics can be
slightly modified giving different dynamical scaling laws. For instance with
{\em node update}, i.e., selecting first a node and then one of its neighbors,
the conservation of the magnetization is not longer fulfilled. Instead the
weighted magnetization is in this case conserved at each time step. With {\em
link update}, where a link is selected at random and then one of its ends is
updated according to the neighbor's spin, the conservation of the
magnetization is restored \cite{Suchecki05b}.

A striking feature of the voter model on several complex networks, including
Small-World, Bar\'abasi-Albert, Erd\H{o}s-R\'enyi, Exponential and Complete
Graph is the lack of complete order in the thermodynamic limit.  In this
article, we provide an analytical insight of the incomplete ordering
phenomenon in heterogeneous networks by studying the evolution and final state
of the system using  a simple mean-field approach. Despite that this approach
is meant  to work well in networks with arbitrary degree distributions but
without node  degree correlations, the qualitative results are rather general
for many networks.  We  obtain analytical predictions for the density of
active links (links connecting nodes with opposite spin) and the mean time to
reach the ordered state as a function of  the system size and the first and
second moments of the degree distribution. These predictions explain numerical
results reported in \cite{Suchecki05b,Castellano05,Suchecki05a}  and they
agree with previous analytical results for ordering times  \cite{Sood05}.

The rest of the article is organized as follows. In section \ref{model}, we
define the model and its updating rule on a general network.  We then develop
in section \ref{MF} a mean-field approach for the time evolution of the
density active links and the link magnetization.  This approximation reveals a
transition at a critical value of the average connectivity $\mu=2$.  When
$\mu$ is smaller than $2$, complete order is reached exponentially fast,
whereas  for $\mu>2$, the system quickly settles in a quasistationary
disordered state  characterized by a constant density of active links whose
value only depends  on $\mu$, independent on the degree distribution.  We find
that $\rho$ is  proportional to the product of the spin densities with a
proportionality  constant that depends on $\mu$.  This relation allows us to
derive an approximate  Fokker-Planck equation for the  magnetization in
section \ref{master}.  This  equation is used in section \ref{approach} to
study the relaxation of a finite  system to the absorbing ordered state and in
section \ref{survival} to obtain  an expression for the survival probability
of independent runs.  The mean time  to reach complete order, calculated in
section \ref{ordering}, shows that the  dependence of the results on the
network topology enters through the first and  the second moments of the
degree distribution only.  Convergence to the ordered state slows down as
$\mu$ approaches $2$, where ordering times seem to  diverge faster than $N$.
The summary and conclusions are provided in section  \ref{summary}.  In the
appendix we present some details of calculations.

\section{The model}
\label{model}

We consider a network composed by a set of $N$ nodes and the links connecting
pair of nodes.  We assume that the network has no degree correlations, i.e.,
the neighbors of each node are randomly selected from the entire set.  We
denote by $P_k$ the degree distribution, which is the fraction of nodes with
$k$ links, subject to the normalization condition $\sum_k P_k = 1$.   In the
initial configuration, spins are assigned the values $1$ or $-1$ with
probabilities given by the initial densities $\sigma_+$ and $\sigma_-$
respectively.  In a single time  step, a node $i$ with spin $S_i$ and one of
its neighbors  $j$ with spin $S_j$ are chosen at random.  Then $i$ adopts
$j$'s spin  ($S_i \to S_i=S_j$) (see Fig.~\ref{update}).  This step is
repeated until the  system reaches complete order and it cannot longer evolve.

\section{Mean -Field theory}
\label{MF}

In order to obtain an insight about the time evolution of the system we
develop a mean-field (MF) approach.  There are two types of links in the
system, links between nodes with different spin or \emph{active links} and
links between nodes with the same spin or \emph{inert links}.  Given that a
single spin-flip update happens only when an active link is chosen, it seems
natural to consider the \emph{global density of active links} $\rho$ as a
parameter that measures the level of activity in the system.

\begin{figure}
\begin{center}
 \vspace*{0.cm}
 \includegraphics[width=0.8\textwidth]{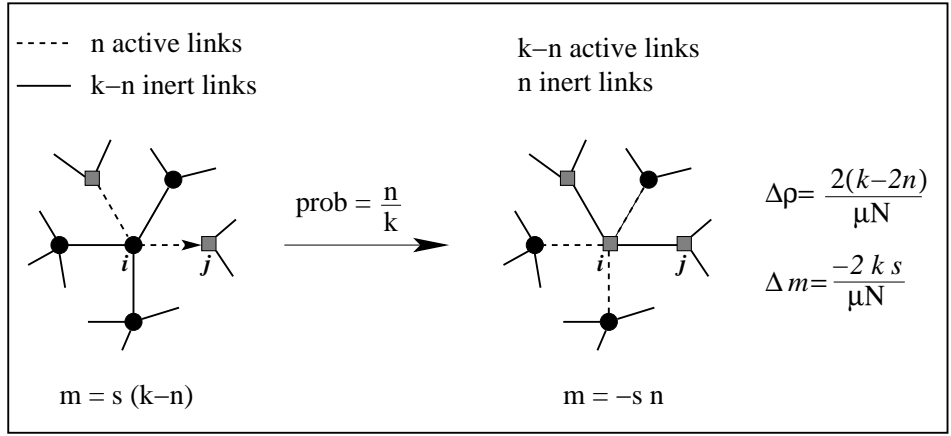}
 \caption{Update event in which a node $i$ with spin $S_i=s$ (black circle) 
   flips its spin to match its neighboring node spin $S_j=-s$ (grey square). 
   The possible values of the spins are $s=\pm1$.  Changes in 
   the density of active links $\rho$ and the link magnetization 
   $m=\rho_{++} - \rho_{--}$ are denoted by $\Delta \rho$ and $\Delta m$
   respectively.} 
 \label{update}
\end{center}
\end{figure} 

In Fig.~\ref{update} we describe the possible changes in $\rho$ and their
probabilities in a time step, when a node $i$ with spin $S_i = s$ ($s=
1$ or $-1$) and degree $k$ is chosen.  We denote  by $n$ the number  of active
links connected to node $i$ before the update.  With probability  $n/k$ an
active link (in this example $i-j$) is randomly chosen.  Node $i$  flips its
state  changing the state of its links from active to inert  and
vice-versa, and  giving a local change of the number of active links 
$\Delta n=k-2n$ and a
global density  change $\Delta \rho = \frac{2(k-2n)}{\mu N}$.  Here $\mu N /2$
is the total  number of links, $\mu \equiv \langle k \rangle = \sum_k k P_k$
is the number  of links per node or average degree.  Assembling these factors,
the change in the  average  density of active links in a single time step of
time interval $dt=1/N$ is  described by the master equation:
\begin{eqnarray}    
\label{drdt1}
\frac{d \rho}{dt} = \sum_{k} P_k \left. \frac{d\rho}{dt} \right|_k 
= \sum_{k} \frac{P_k}{1/N} \sum_{n=0}^k B(n,k)\, \frac{n}{k}\,  
\frac{2 (k-2n)}{\mu N}, 
\end{eqnarray}
where $B(n,k)$ is the probability that $n$ active links are connected to a
node of degree $k$,  and $ \left. \frac{d\rho}{dt} \right|_k $ denotes the 
average change in $\rho$ when a node of degree $k$ is chosen.   Given that,
during the evolution, the densities of $+$ and $-$ spins are not the  same, we
expect that $B(n,k)$ will depend on the spin of node $i$.  For instance, when
the system is about to reach the $+$ fully ordered state, we expect a
configuration where most of the neighbors of a given node (independent on its
spin) have $+$ spin, thus the probability that a link connected to a node with
spin $+$ ($-$) is active will be close to zero (one).  Therefore, we take 
$B(n,k)$ as the average probability over the two types of spins 
\begin{equation}
B(n,k) = \sum_{s=\pm} \sigma_s \; B(n,k|s),
\label{Bnk}
\end{equation}
where $B(n,k|s)$ is the conditional probability that $n$ of the $k$ links
connected to a node are active, given that the node has spin $s$. 
Replacing Eq.~(\ref{Bnk}) into Eq.~(\ref{drdt1}) we obtain  
\begin{eqnarray}
\frac{d \rho}{dt} &=& \frac{2}{\mu} \sum_{k} P_k \sum_{s=\pm} \sigma_s 
\sum_{n=0}^k B(n,k|s) \;\frac{n}{k}\; (k-2n) \\
\label{drdt2} 
&=& \frac{2}{\mu} \sum_{k} P_k \sum_{s=\pm} \sigma_s \left[ 
\langle n \rangle_{k,s} - \frac{2}{k} \langle n^2 \rangle_{k,s} \right], 
\end{eqnarray}
where $\langle n \rangle_{k,s}$ and $\langle n^2 \rangle_{k,s}$ are the first
and the second moments of $B(n,k|s)$ respectively.  

In order to calculate $B(n,k|s)$ we assume that only correlations between  the
states of first neighbors are relevant, neglecting second or higher  neighbors
correlations.  Therefore, we consider the conditional probability  $P(-s|s)$,
that a neighbor of node $i$ has spin $-s$ given that $i$ has  spin  $s$, to be
independent of the other neighbors of $i$.  This is known in  the lattice
models literature  with the name of \emph {pair approximation}, and it is
supposed to work only  in networks without degree correlations.  Thus,
$B(n,k|s)$ becomes the  binomial  distribution with $P(-s|s)$ as the single
event probability that a link connected to $i$ is active.  $P(-s|s)$ can be
calculated as the average fraction of neighbors with spin $-s$ to a node with
spin $s$, i.e., the ratio between the total number $\rho \,\mu N/2$ of  $s \to
-s$ links and the total number $\mu \,\sigma_s N$ of links connected to nodes
with spin  $s$.  We have used the symmetry in the states of the voter model
and assumed that the average degrees of nodes holding spin $1$ and $-1$ are the
same and  equal to $\mu$.  We have numerically checked that the last is valid
for the original voter model, but if the two states are not equivalent or a
biased is introduced, the average degrees are different. Then,  $P(-s|s)= \rho
/ 2\, \sigma_s$, and the  first and the second moments of $B(n,k|s)$ are
\begin{eqnarray*}
\langle n \rangle_{k,s} &=& \frac{k \rho}{2 \sigma_s} \\
\langle n^2 \rangle_{k,s} &=& 
\frac{k \rho}{2\sigma_s} + \frac{k(k-1) \rho^2}{4 \sigma_s^2}.
\end{eqnarray*} 
Replacing these expressions for the moments in Eq.~(\ref{drdt2}) and
performing the sums we finally obtain
\begin{eqnarray}
\label{drdt3}
\frac{d \rho}{dt} = \frac{2 \rho}{\mu} \left[ (\mu-1) 
  \left( 1- \frac{\rho}{2 \sigma_+ (1-\sigma_+)} \right) - 1 \right].
\end{eqnarray}
Equation (\ref{drdt3}) is the master equation for the time evolution of $\rho$
as a function of the spin density $\sigma_+(t)$.  It has two stationary
solutions, but depending on the value of $\mu$, only one is stable.  For 
$\mu \leq 2$, the stable solution $\rho=0$ corresponds to a fully ordered 
frozen system.  For $\mu >2$, the stable solution is    
\begin{equation}
\rho(t) = 4 \,\xi(\mu)\, \sigma_+(t) \left[1-\sigma_+(t)\right],
\label{r-t1} 
\end{equation}
where we define 
\begin{equation}
\xi (\mu) \equiv \frac{(\mu-2)}{2(\mu-1)},
\label{xi}
\end{equation}
corresponding to a partially ordered system, composed by a fraction $\rho > 0$
of active links, as long as $\sigma_+ \not= 0,1$. 

\begin{figure}
\begin{center}
 \vspace*{0.cm}
 \includegraphics[width=0.75\textwidth]{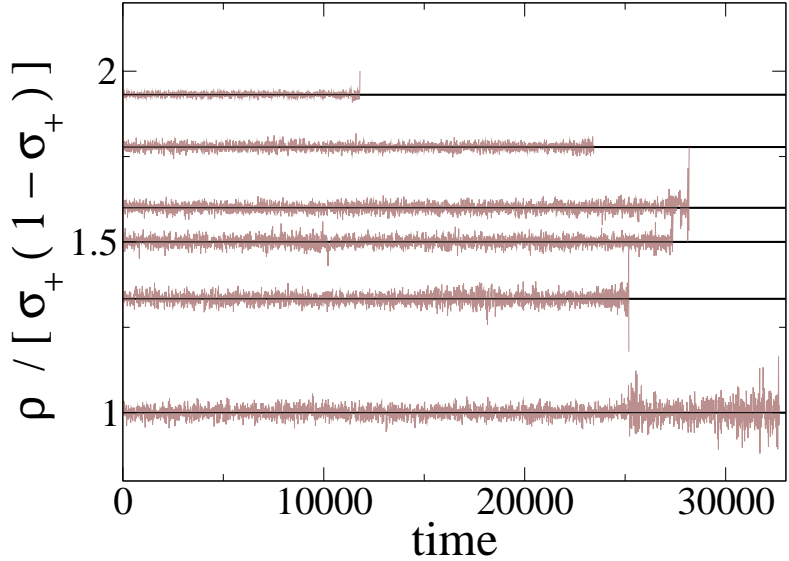}
 \caption{Ratio between the density of active links and the product of the
   spin densities vs time in one realization of the voter model dynamics on 
   networks with $N=10000$ nodes and values of $\mu=3,4,5,6,10$ and $30$ 
   (bottom to top).  Solid horizontal lines are the constant values 
   $4 \, \xi= \frac{2(\mu-2)}{(\mu-1)}$.}
 \label{ratio}
\end{center}
\end{figure} 

In Fig.~\ref{ratio} we test Eq.~(\ref{r-t1}) by plotting the time evolution of
the ratio between $\rho$ and $\sigma_+ (1-\sigma_+)$ in a single realization, 
for various values of $\mu$.  We observe that, even though the ratio varies
over time, it fluctuates around  the constant value $4\, \xi$ predicted by
Eq.~(\ref{r-t1}).  It is worth noting that the behavior of the ratio is the
same from times of order one to the end of the realization, where fluctuations
increase in amplitude before the system reaches complete order.  We also notice
that  fluctuations decrease as $\mu$ increases, and  they become zero in the
complete graph case ($\mu = N-1$), where we have 
$\rho(t)= 2\, \sigma_+(t)\left[1-\sigma_+(t)\right]$, for $N \gg 1$.

In infinite large systems, fluctuations in $\sigma_+(t)$ vanish.  Therefore,
in a single realization we would see that $\sigma_+(t) = \sigma_+(0)$ for all
$t>0$ and that $\rho$ reaches an infinite long lived stationary state with
$\rho^s= 4 \,\xi \sigma_+(0)\left[1-\sigma_+(0)\right]$.  Then, for networks 
with average degree $\mu>2$,  full order is never reached in the thermodynamic 
limit.  

In finite size networks, fluctuations  eventually drive the system to one of
the two absorbing states, $\sigma_+=1$  or $\sigma_+=0$, characterized by the
absence of active links ($\rho=0$).  Although the parameter $\rho$ is useful 
for finding an absorbing state, it does not allow us to know which of
the two states is reached.  For this reason we introduce  the \emph{link
magnetization} $m = \rho_{++} - \rho_{--}$, where $\rho_{++}$ ($\rho_{--}$)
are the density of links connecting two nodes with spins $1$  ($-1$).  It
measures the level of  order in the system, $m=1$ ($m=-1$) corresponding to
the $+$ ($-$) fully ordered absorbing state and $m=0$ representing the totally
mixed disordered state.  Given that  $\rho$ becomes zero when $m$ takes the
values $\pm 1$, we guess that $\rho$  should be proportional to $1-m^2$.  To
prove this, we first relate $\sigma_s$ with $\rho_{ss}$ by calculating the 
total number of links coming out from nodes with spin $s$.  This number of 
links is $\mu \sigma_s N$, from which $\rho \mu N/2$ are $s \to -s$
links, and $\rho_{ss} \mu\, N$ are $s \to s$ links.  We arrive to
\begin{eqnarray}
\label{r++}
\rho_{ss}&=&\sigma_{s}-\rho/2. \nonumber
\end{eqnarray}
Then, the link magnetization is simply the spin magnetization
\begin{equation}
m = \rho_{++} - \rho_{--} = \sigma_+ - \sigma_- = 2 \sigma_+ -1. 
\label{m-sigma}
\end{equation}
Combining Eqs.~(\ref{r-t1}) and (\ref{m-sigma}) we obtain that, neglecting 
fluctuations, $\rho$ and $m$ are related through the equation
\begin{equation}
\rho(t) = \xi \left[1-m^2(t)\right].
\label{r-mag}
\end{equation} 

\begin{figure}
\begin{center}
 \vspace*{0.cm}
 \includegraphics[width=0.8\textwidth]{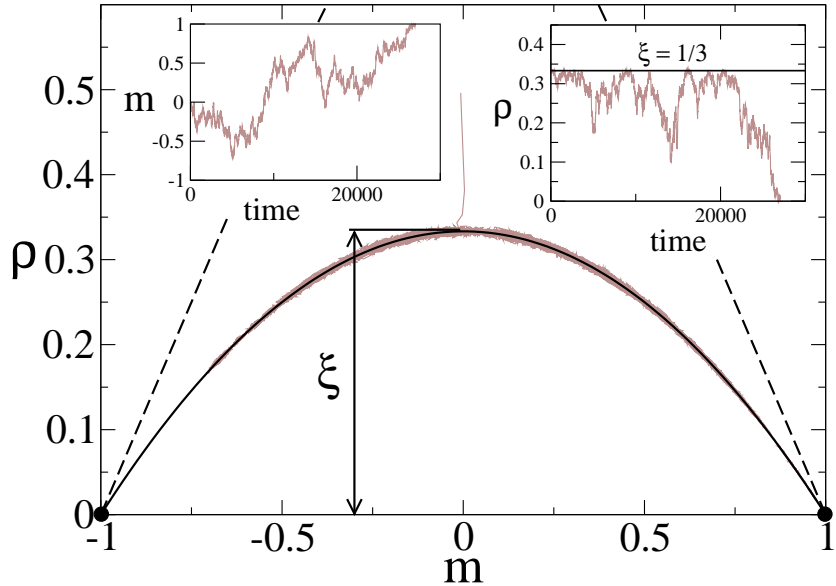}
 \caption{Trajectory of the system in a single realization plotted on the
   active links-link magnetization ($\rho-m$) plane, for a network of size
   $N=10^4$ and average degree $\mu=4$.  Insets: Time  evolution of $m$ (left)
   and $\rho$ (right) for the same realization.  We note that $\rho$ and $m$ 
   are not independent but fluctuate in coupled  manner, following a parabolic
   trajectory described by $\rho=\frac{1}{3}(1-m^2)$ from Eq.(\ref{r-mag})
   (solid line).}
 \label{r-m}
\end{center}
\end{figure} 

Fig.~\ref{r-m} shows $\rho$ vs $m$ in one realization with $\mu=4$ and
$N=10^4$.  The system starts with equal density of $+$ and $-$
spins ($m=0$ and $\rho=1/2$), and after an initial transient of order one, in
which $m$ stays close to zero and $\rho$ decays to a value similar to $\xi$,  
$\rho$ fluctuates around the parabola described by Eq.~(\ref{r-mag}).  This
particular trajectory ends at the $(m=1,\rho=0$) absorbing state.

\section{Master Equation for the link magnetization}
\label{master}

In order to study the time evolution of the system we start by deriving a
master equation for the probability $P(m,t)$ that the system has link
magnetization $m$ at time $t$.  In a time step, a node with spin $s$ and
degree $k$ flips its spin with probability  $\sigma_s P(-s|s) = \rho/2$, after 
which
the magnetization changes by  $\Delta m = s\,\delta_k$, with
$\delta_k=\frac{2k}{\mu N}$ (see   Fig.~\ref{update}), and with probability
$\sigma_s \left[1-P(-s|s)\right] = \sigma_s (1-\rho/2 \sigma_s)$ its spin
remains  unchanged.  We have used that the density  of $s$ spins and the
conditional  probability $P(-s|s)$ in the subset of nodes with degree $k$ is
independent on  $k$ and equal to the global density $\sigma_s$ (this was first
noticed in  \cite{Sood05} and \cite{Suchecki05a}).  Using Eq.~(\ref{r-mag}) we
can write  the probabilities of the possible changes in $m$ due to the
selection of a  node of degree $k$ as
\begin{eqnarray}
W_{m \to m- \delta_k} &=& \frac{\xi}{2} \left(1-m^2 \right) P_k \nonumber \\
\label{W}
W_{m \to m+ \delta_k} &=& \frac{\xi}{2} \left(1-m^2 \right) P_k \\
W_{m \to m } &=& \left[ 1- \xi \left(1-m^2 \right) \right] P_k. \nonumber
\end{eqnarray}
Thus, the problem is reduced to the motion of a symmetric random walk in the 
$(-1,1)$ interval, with absorbing boundaries at the ends and hopping 
distances and their probabilities that depend on the walker's position $m$ and 
the degree distribution $P_k$.  The time evolution of $P(m,t)$ is described 
by the master equation 
\begin{eqnarray}
\label{P-m-t}
P(m,t+\delta t) &=& \sum_k  P_k \Biggl\{ W_{m+\delta_k \to m}\;
P \left(m+\delta_k,t \right)+
W_{m-\delta_k \to m} \; P \left(m-\delta_k,t \right)  \nonumber \\
&+& W_{m \to m} \; P(m,t) \Biggl\}  \nonumber \\ 
&=&\sum_k  P_k \Biggl\{ \frac{\xi}{2} 
\left[1-\left(m+ \delta_k\right)^2\right] 
P \left(m+\delta_k,t \right)  \\ 
&+& \frac{\xi}{2} \left[1-\left( m-\delta_k \right)^2\right] 
P \left(m-\delta_k,t \right) + \left[1-\xi(1-m^2) \right]  P(m,t) \Biggl\}, 
\nonumber
\end{eqnarray}
where $\delta t=1/N$ is the time step corresponding to a spin-flip attempt.
In Eq.~(\ref{P-m-t}), the probability that the walker is at site $m$ at time
$t+\delta t$ is  written as the sum of the probabilities for all possible
events that take the  walker from a site $m + \Delta$ to site $m$, with
$\Delta=0,\pm \delta_k$ and $k \geq 0$.  The probability of a single event is
the probability $P(m + \Delta,t)$ of being at site $m + \Delta$ at time $t$
times the probability  $W_{m+\Delta \to m}$ of hopping to site $m$.  Expanding
Eq.~({\ref{P-m-t}) to second order in $m$ and  first order in $t$ we obtain 
\begin{eqnarray*}
N \delta t \frac{\partial P}{\partial t} = \frac{2 \, \xi}{\mu^2 N} 
\sum_k P_k \, k^2 \Biggl\{
- 2 P - 4\, m \,\frac{\partial P}{\partial m} + (1-m^2) \,
\frac{\partial^2 P}{\partial m^2} \Biggr\}.
\end{eqnarray*}
Thus, in the continuum limit ($\delta t=1/N \to 0$ as $N \to \infty$), we
arrive to the Fokker-Planck equation 
\begin{equation}
\frac{\partial P(m,t')}{\partial t'} = \frac{\partial^2}{\partial m^2} 
\left[ (1-m^2) P(m,t') \right],
\label{dPdt}
\end{equation} 
where $t' \equiv t / \tau$ is a rescaled time, 
\begin{equation}
 \tau \equiv \frac{\mu^2 N}{2 \,\xi(\mu) \,\mu_2} = 
 \frac{(\mu-1) \mu^2 N}{(\mu-2) \,\mu_2}
 \label{tau}
\end{equation}
is an intrinsic time scale of the system and $\mu_2 = \sum_k k^2 P_k$ is the
second moment of the degree distribution.  We shall see in section 
\ref{ordering}
that the time to reach the ordered state equals $\tau$ times a function of the
initial magnetization.  Note that, in complete graph, the corresponding
Fokker-Planck equation derived for instance in \cite{Ben-Avraham90},  has the
same form as Eq.~(\ref{dPdt}) with $t'=t/N$, obtained as a particular case of
a graph with distribution $P_k=\delta_{k,\mu}$, $\mu=N-1$
and $\mu_2=\mu^2$.  The general solution to Eq.~(\ref{dPdt}) is given by the
series expansion \cite{Ben-Avraham90,Slanina04}
\begin{equation} 
P(m,t')=\sum_{l=0}^\infty A_l \;C_l^{3/2}(m)\;e^{-(l+1)(l+2)\,t'},
\label{P-t}
\end{equation}  where $A_l$ are coefficients determined by the initial
condition and $C_l^{3/2}(x)$ are the Gegenbauer polynomials \cite{Grandshteyn}
page 980.
Equation (\ref{P-t}) is of fundamental importance because it allows to find
the two most relevant magnitudes in the voter model dynamics, namely, the 
average density of active links and the survival probability, as we shall see 
in sections \ref{approach} and \ref{survival} respectively.

\section{Approach to the final frozen state}
\label{approach}

We are interested in how the average density of active links 
$\langle \rho \rangle$ decays to zero, where
$\langle \cdot \rangle$ denotes an average over many independent realizations 
of the dynamics starting from the same initial spin densities.  Using 
Eq.~(\ref{r-mag}) we can write 
\begin{eqnarray}
\label{integral}
\langle \rho(t') \rangle = \xi \langle 1-m^2(t') \rangle 
= \xi \int_{-1}^1 dm \;(1-m^2)\; P(m,t'),
\end{eqnarray}
with $P(m,t')$ given by Eq.~(\ref{P-t}).  The solution to the above integral 
with an initial magnetization $m_0 = 2 \sigma_+(0)-1$ is (see appendix 
\ref{ave-den})
\begin{equation}
\label{rho-ave1}
\langle \rho(t') \rangle = \xi (1-m_0^2) \;e^{-2 t'}, 
\end{equation}
and replacing back $t'$ and $\xi(\mu)$ we finally obtain 
\begin{equation}
\langle \rho(t) \rangle=\frac{(\mu-2)}{2(\mu-1)} (1-m_0^2) e^{-2 \,t/\tau}.
\label{r-t2}
\end{equation}
We find that for $\mu>2$, $\langle \rho(t) \rangle$ has an exponential decay 
with a time constant $\tau/2$, whose inverse gives the rate at which 
$\langle \rho \rangle$ decays.  Given that $\tau$ is proportional to 
$N$ (Eq.~(\ref{tau})), the decay becomes slower for increasing system sizes.  
Eventually, in the limit of an infinite large network 
$\langle \rho(t) \rangle$ remains at the constant value $\xi(1-m_0^2)$ as it 
was discussed in section \ref{MF}, while in a finite network, 
$\langle \rho(t) \rangle$ reaches zero in a time of order $\tau$.    

We have simulated the voter model on various types of random networks: 
degree-regular random graph (DR), Erd\H{o}s-R\'enyi graph (ER), Exponential 
network (EN) and Bar\'abasi-Albert network (BA).
\ignore{and complete graph (CG).}  In 
Fig.~\ref{r-t-NN} we observe that the analytical prediction 
(Eq.~(\ref{r-t2})) is in good agreement with numerical simulations on these
four networks.  
\begin{figure}[t]
\begin{center}$
\begin{array}{cc}
\includegraphics[width=3.0in]{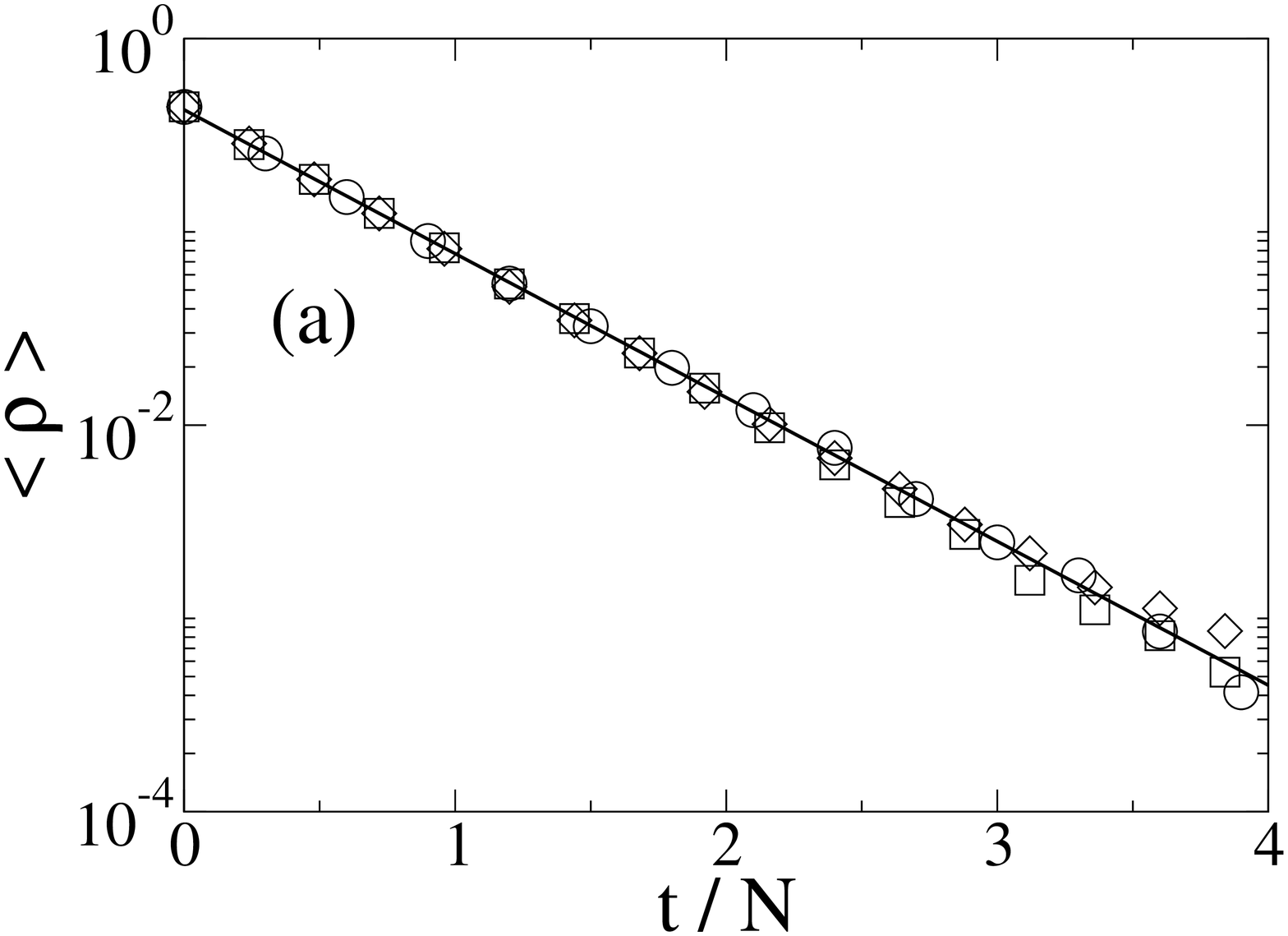} &
\includegraphics[width=3.0in]{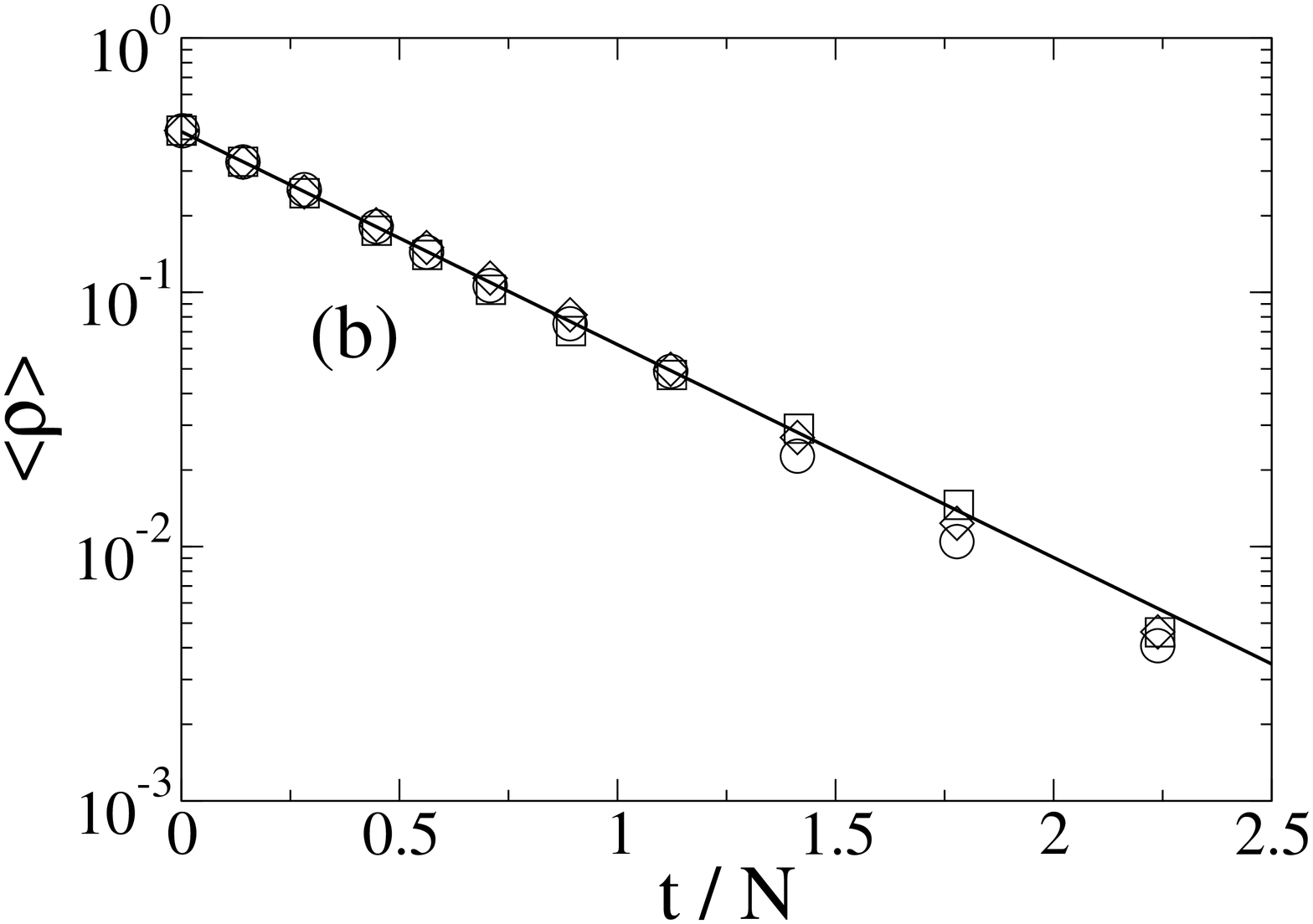} \\
\includegraphics[width=3.0in]{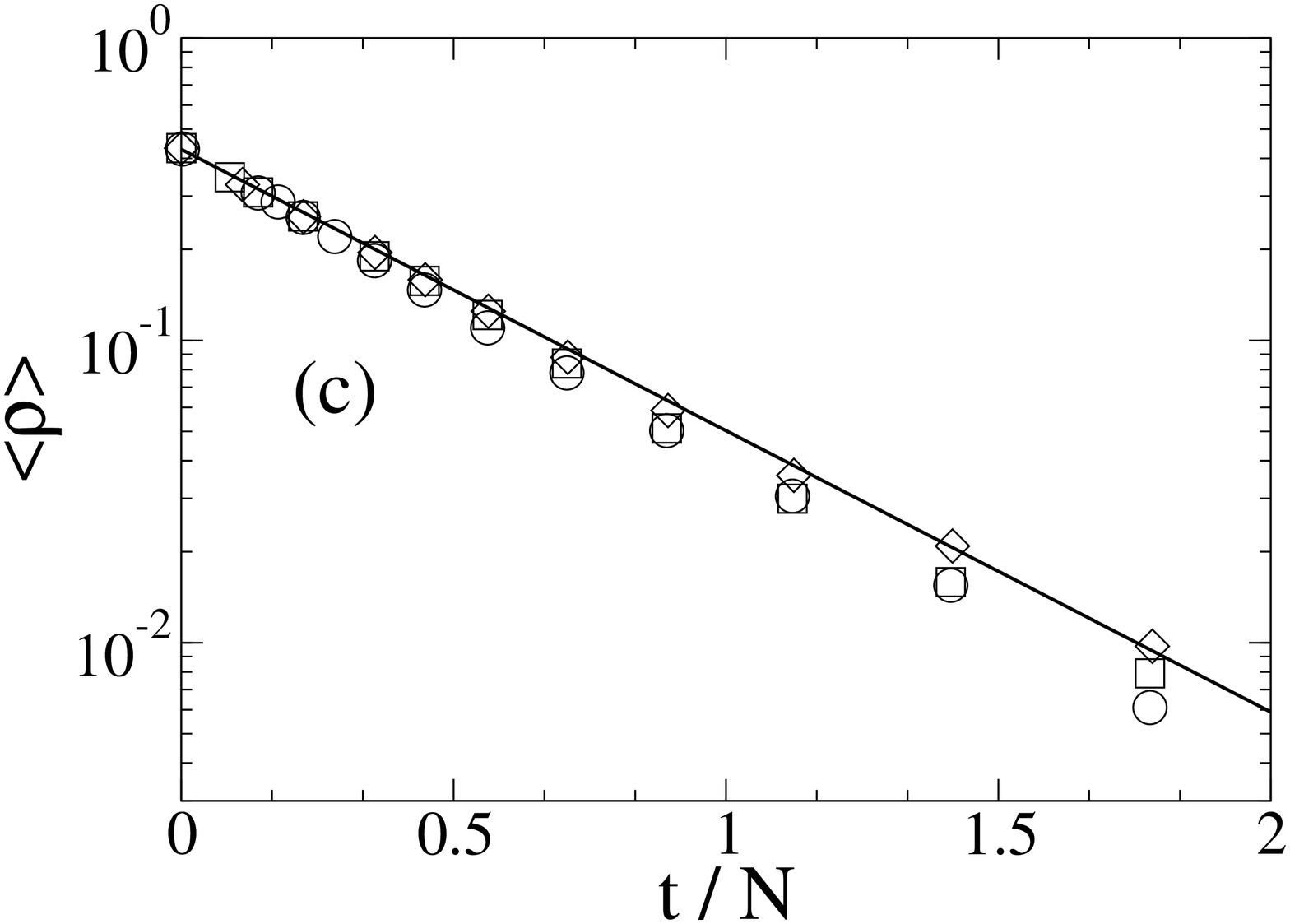} &
\includegraphics[width=3.0in]{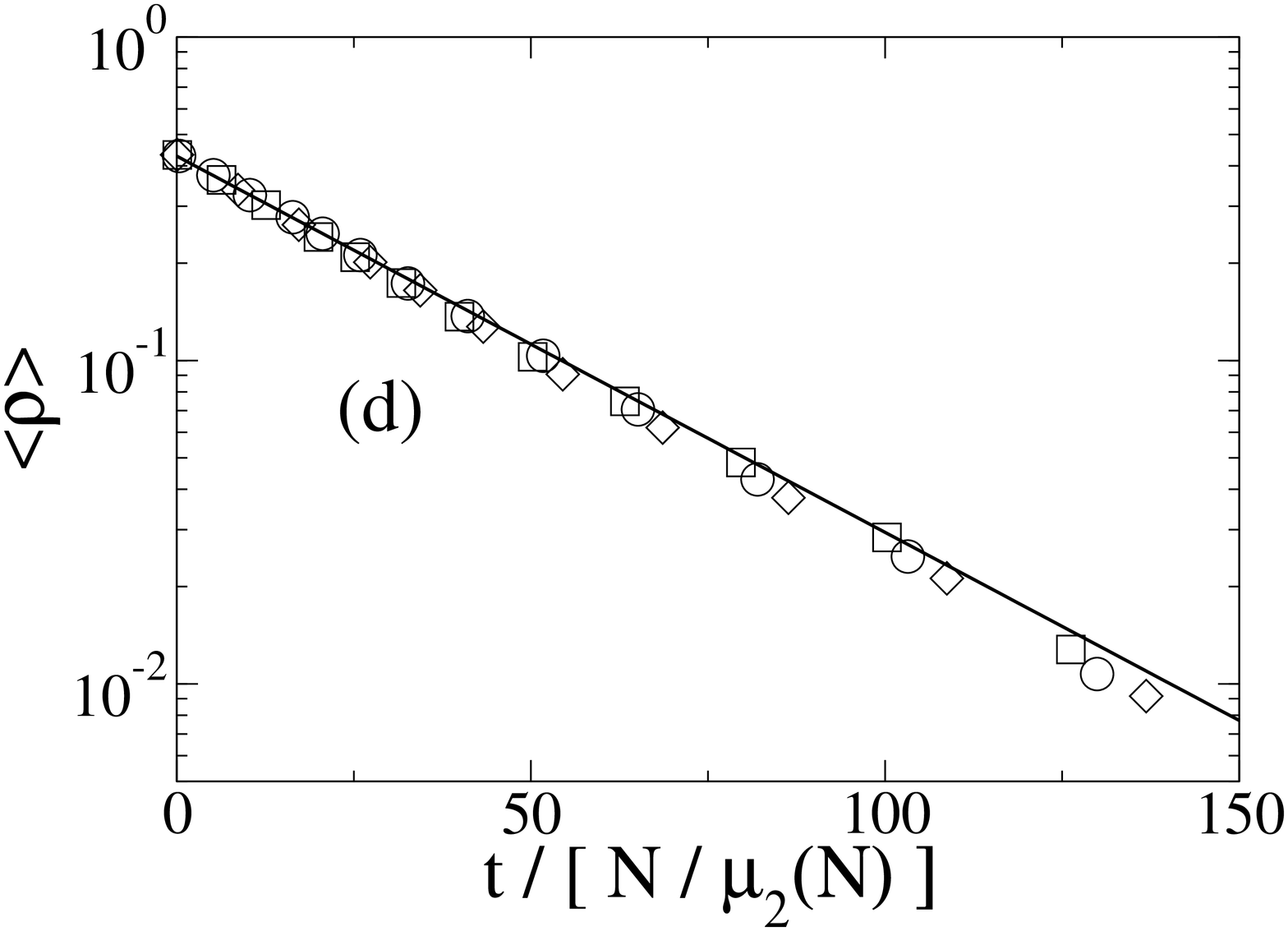} 
\end{array}$
\end{center}
\caption{Time evolution of the average density of active links 
  $\langle \rho(t) \rangle$ for (a) DR, (b) ER, (c) EN and (d) BA networks 
  with average degree $\mu=8$.  The open symbols correspond to 
  networks of different sizes: $N=1000$ (circles),  
  $N=5000$ (squares) and $N=10000$ (diamonds).  Solid lines are the
  analytical predictions from Eq.~(\ref{r-t2}).  The average was taken over 
  $1000$ independent realizations, starting from a uniform distribution with 
  magnetization $m_0=0$.}
\label{r-t-NN}
\end{figure}
For a fix average degree $\mu$ and system size $N$, $\tau$ is determined by
the second moment $\mu_2$ of the network degree distribution $P_k$.  For
these particular networks, $\mu_2$ can be written as a  function of $\mu$,
because $P_k$ only depends on $\mu$ and $k$.  As a  consequence of this,
$\tau(\mu,N)$ is only a function of $\mu$ and $N$.   The values of $\tau$ and
$\mu_2$ in the large $N$ limit are summarized in table \ref{table}.  For the
case of DR, ER and EN, $\langle \rho \rangle$ is a function of $t/N$ as it is
shown in Fig.~\ref{r-t-NN} and $\mu_2$ is finite and  independent on $N$. We
have checked that the scaling  works very well for networks of size $N > 100$.
For BA networks, $\mu_2$  diverges with $N$ (see calculation details in
appendix \ref{mu2-tau}), thus we rescaled the x-axis by  $N/\mu_2(N)$ in order
to obtain an overlap for the curves of different system  sizes.

\begin{table}[h]
\begin{center}
\begin{tabular}{|c|c|c|c|}
  \hline
  Network&$P_k$& $\mu_2$& $\tau(\mu,N)$ \\
  \hline
  DR&$\delta_{k,\mu}$&$\mu^2$&$\frac{(\mu-1)}{(\mu-2)}N$ \\
  \hline 
  ER&$e^{-\mu}\frac{\mu^k}{k!}$&$\mu(\mu+1)$&$\frac{\mu(\mu-1)}
  {(\mu+1)(\mu-2)}N$ \\
  \hline
  EN&$\frac{2\,e}{\mu}\exp\left({-\frac{2 k}{\mu}}\right)$&$\frac{5}{4}\mu^2$&
  $\frac{4(\mu-1)}{5(\mu-2)}N$ \\
  \hline 
  BA&$\frac{\mu(\mu+2)}{2 k (k+1) (k+2)}$&
  $\frac{\mu(\mu+2)}{4}\ln\left(\frac{\mu(\mu+2)^3\,N}{(\mu+4)^4}\right)$&
  $\frac{4\mu(\mu-1)N/(\mu^2-4)}
  {\ln\left(\frac{\mu(\mu+2)^3}{(\mu+4)^4}N\right)}$ \\
  \hline 
  CG&$\delta_{k,N-1}$&$(N-1)^2$&$N$ \\
  \hline 
\end{tabular}
\end{center}
\caption{Node degree distribution $P_k$, its second moment $\mu_2$ and the
  decay time constant of the average density of active links $\tau$, for 
  different networks.}
\label{table}
\end{table}

\section{Survival probability}
\label{survival}

In the last section we found that the density of active links, when averaged 
over many runs, decays exponentially fast to zero.  In
doing this average at a particular time $t$, we are considering all runs, even 
those that die before $t$ and, therefore, contribute with $\rho=0$ to the
average. In order to gain an insight about the evolution of a single run 
\cite{Castellano05}, we consider the density of active links
averaged only over surviving runs 
$\langle \rho^{\mbox{\scriptsize surv}}(t)\rangle$.  If we
define the survival probability $S(t)$ as the probability that the system has 
not reached the fully ordered state up to time $t$, then we can write 
$\langle\rho(t)\rangle = S(t)\langle\rho^{\mbox{\scriptsize surv}}(t)\rangle$.

In the $1d$ random walk mapping that we discussed in section \ref{master}, 
$S(t)$ 
corresponds to the probability that the RW is still alive at time $t$, that is
to say, that it has not hit the absorbing boundaries $m=\pm 1$ up to time
$t$.  If at time $t=0$, we launch many walkers from the same position
$m_0$, each of which representing an individual run, then $S(t)$ can be
calculated as the fraction of surviving walkers at time $t$
\begin{equation}
\label{S-t2}
  S(t)=\int_{-1}^1 dm \; P(m,t).
\end{equation}
The result of this integral for symmetric initial conditions ($m_0=0$) is
given by the series (see appendix \ref{surv})
\begin{equation}
S(t)=\sum_{l=0}^{\infty} \frac{(-1)^l (4l+3) (2l-1)!!}{(2l+2)!!} 
\exp \left(-\frac{2 (2l+1)(l+1)\;t}{\tau(\mu,N)}\right).
\label{S-t0}
\end{equation}

\begin{figure}
\begin{center}
 \vspace*{0.cm}
 \includegraphics[width=0.8\textwidth]{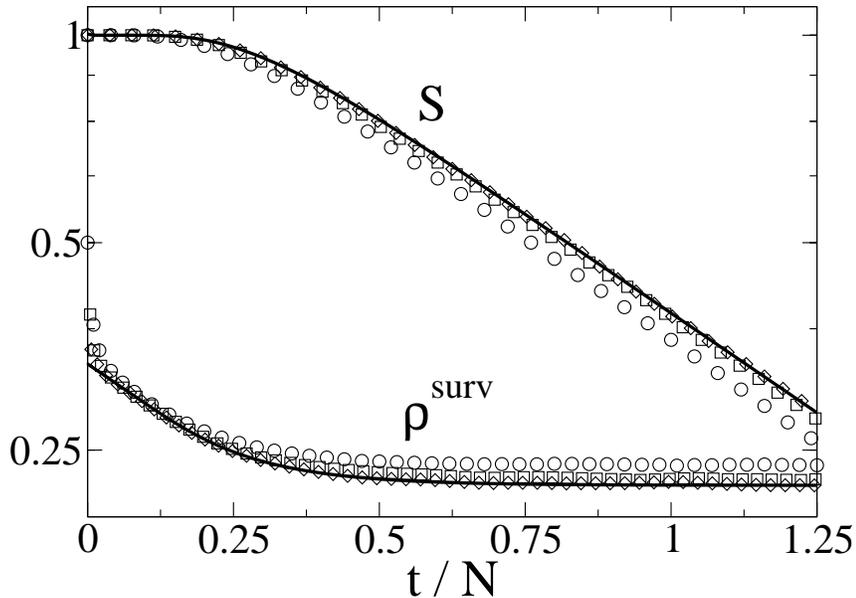}
\end{center}
 \caption{Survival probability $S$ and average density of active links in 
   surviving runs $\langle \rho^{\mbox{\scriptsize surv}} \rangle$ vs the
   rescaled time $t/N$ for DR networks with degree $\mu=4$ and sizes $N=100$ 
   (circles), $N=400$ (squares) and $N=1600$ (diamonds).  Top and bottom solid 
   lines are the analytical solutions $S(t)$ and 
   $\langle \rho^{\mbox{\scriptsize surv}}\rangle=\langle \rho(t)\rangle /S(t)$
   respectively obtained using equations (\ref{S-t0}) and (\ref{r-t2}).} 
 \label{Sna}
\end{figure} 

As we observe in Fig.~\ref{Sna} there are two regimes.  For  $t \ll N$, is
$S(t) \simeq 1$. For $t \gtrsim N/4$, only the first term corresponding to 
the
lowest $l$ ($l=0$) gives a significant contribution to the series, thus
neglecting the  terms with $l>0$ gives  $S(t) \simeq \frac{3}{2} \exp
\left(-\frac{t}{\tau(\mu,N)}\right)$.   For a general initial condition $m_0$,
we obtain that the survival probability decays as
\begin{equation}
S(t) \simeq \frac{3}{2} (1-m_0^2) 
\exp\left(-\frac{2 (\mu-2) \,\mu_2}{(\mu-1)\mu^2} \frac{t}{N} \right)
~~~\mbox{for}~~~t > N.
\label{S-t}
\end{equation}
Using Eqs.~(\ref{r-t2}) and (\ref{S-t}) we finally obtain that the density of 
active links in surviving runs is
\begin{eqnarray}
\langle \rho^{\mbox{\scriptsize surv}}(t) \rangle \simeq 
\cases{
\frac{(\mu-2)}{2(\mu-1)} (1-m_0^2) e^{-2\, t / \tau} &
for $t \ll N$;\\
\frac{(\mu-2)}{3(\mu-1)} & for $t \geq N$.\\}
\label{r-ave}
\end{eqnarray}
We find that the system reaches in a time of order $N$ a partially ordered 
steady state, in which the average density of active links is 
\begin{equation}
\label{xi5}
\frac{2}{3} \xi(\mu) = \frac{(\mu-2)}{3(\mu-1)}.
\end{equation}
In fig.~\ref{plateau}
we plot the average height of the  plateau as a function of $\mu$ obtained
from numerical simulations on a Bar\'abasi-Albert network and a degree-regular
random graph.  As Eq.~(\ref{xi5}) shows, the average
plateau value $2 \, \xi/3$ is only a function of the first moment of the
distribution, as long as the network is random.  The plateau is also 
independent on
the initial condition $m_0$, and the system size $N$ for $N$ large.  

\begin{figure}
\begin{center}
 \vspace*{0.cm}
 \includegraphics[width=0.8\textwidth]{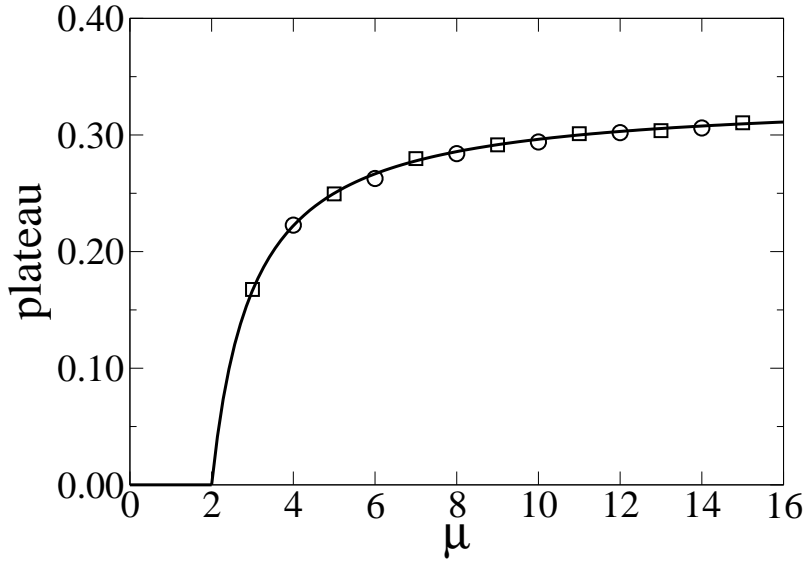}
 \caption{Average height of the plateau for BA (circles) and DR (squares) 
   networks of size $N=10000$.  The solid line is the analytical prediction
   $\frac{(\mu-2)}{3(\mu-1)}$.}
 \label{plateau}
\end{center}
\end{figure} 

A natural question is about the typical size of \emph{spin domains} in the 
stationary state, where we use the term domain to identify a set of connected 
nodes with the same spin.  Numerical simulations reveal that the system is 
always composed by two large domains with opposite spin until by fluctuations 
one of them takes over and the system freezes.  This can be explained using 
percolation transition arguments on random graphs.  Two connected
nodes belong to the same domain if the link that connects them is inert, and
this happens with probability $q=1-\rho$.  Then, a domain that spans the system
exists if $q > q_c = \frac{1}{\kappa-1}$, with 
$\kappa=\frac{\mu_2}{\mu}$ \cite{Cohen00}.  This gives a critical density 
\begin{equation}
\rho_c = \frac{\mu_2-2\mu}{\mu_2-\mu}.     
\end{equation}
Given that $\mu_2 \geq \mu^2$, we have 
$\rho_c \geq \frac{\mu-2}{\mu-1}=2 \,\xi$, and because the density of active 
links in one realization is equal or smaller than $\xi$ (see Fig.~\ref{r-m}), 
the system remains in the ``percolated phase'', i.e., most of the nodes with 
the same spin are connected forming a giant domain of the order of the system 
size.

\section{Ordering time in finite systems}
\label{ordering}

A quantity of interest in the study of the voter model is the mean time to
reach the fully ordered state when initially the system has magnetization
$m$.  In the random walk terminology of section \ref{master}, this is 
equivalent to the 
mean exit time $T(m)$, i.e., the time that the walker takes to reach 
either absorbing boundary $m=\pm 1$ by the first time, starting from the 
position $m$.  $T(m)$ obeys the following recursion formula:
\begin{eqnarray*}
 T(m) &=&\sum_k  P_k \Biggl\{ 
 \frac{\xi}{2} (1-m^2) \left[T(m+\delta_k)+ \delta t \right]  \\
 &+&\frac{\xi}{2} (1-m^2) \left[T(m-\delta_k)+ \delta t \right] 
 + \left[ 1-\xi(1-m^2) \right] \left[T(m)+ \delta t \right] \Biggl\}, 
\end{eqnarray*}
with boundary conditions
\begin{equation}
T(-1)=T(1)=0.
\end{equation}
The mean exit time starting from site $m$ equals the probability of taking a 
step to a site $m+\Delta$ times the exit time starting from this site.  We
then have to sum over all possible steps $\Delta=0,\pm \delta_k$ and add the
time interval $\delta t$ of a single step.  In the continuum limit 
($\delta_k, \delta t \to 0$ as $N \to \infty$) this equation becomes 
\begin{equation}
\frac{d^2 T(m)}{dm^2} = - \frac{\tau}{(1-m^2)},
\end{equation}
where $\tau$ is defined in Eq.~(\ref{tau}).  The solution to this equation is 
\begin{eqnarray*}
  T(m) = \tau \left[\frac{1+m}{2} \ln \left(\frac{1+m}{2} \right) + 
                   \frac{1-m}{2} \ln \left(\frac{1-m}{2} \right) \right],
\end{eqnarray*}
or, in terms of the initial density of $+$ spins $\sigma_+ = (1+m)/2$ 
\begin{equation}
  T(\sigma_+) = -  \frac{(\mu-1) \mu^2}{(\mu-2) \,\mu_2} \; N \; 
  \left[\sigma_+ \ln \sigma_+ + (1-\sigma_+) 
    \ln \left(1-\sigma_+\right) \right].                   
\label{tau-sigma}
\end{equation}

\begin{figure}[t]
\begin{center}$
\begin{array}{cc}
\includegraphics[width=3.0in]{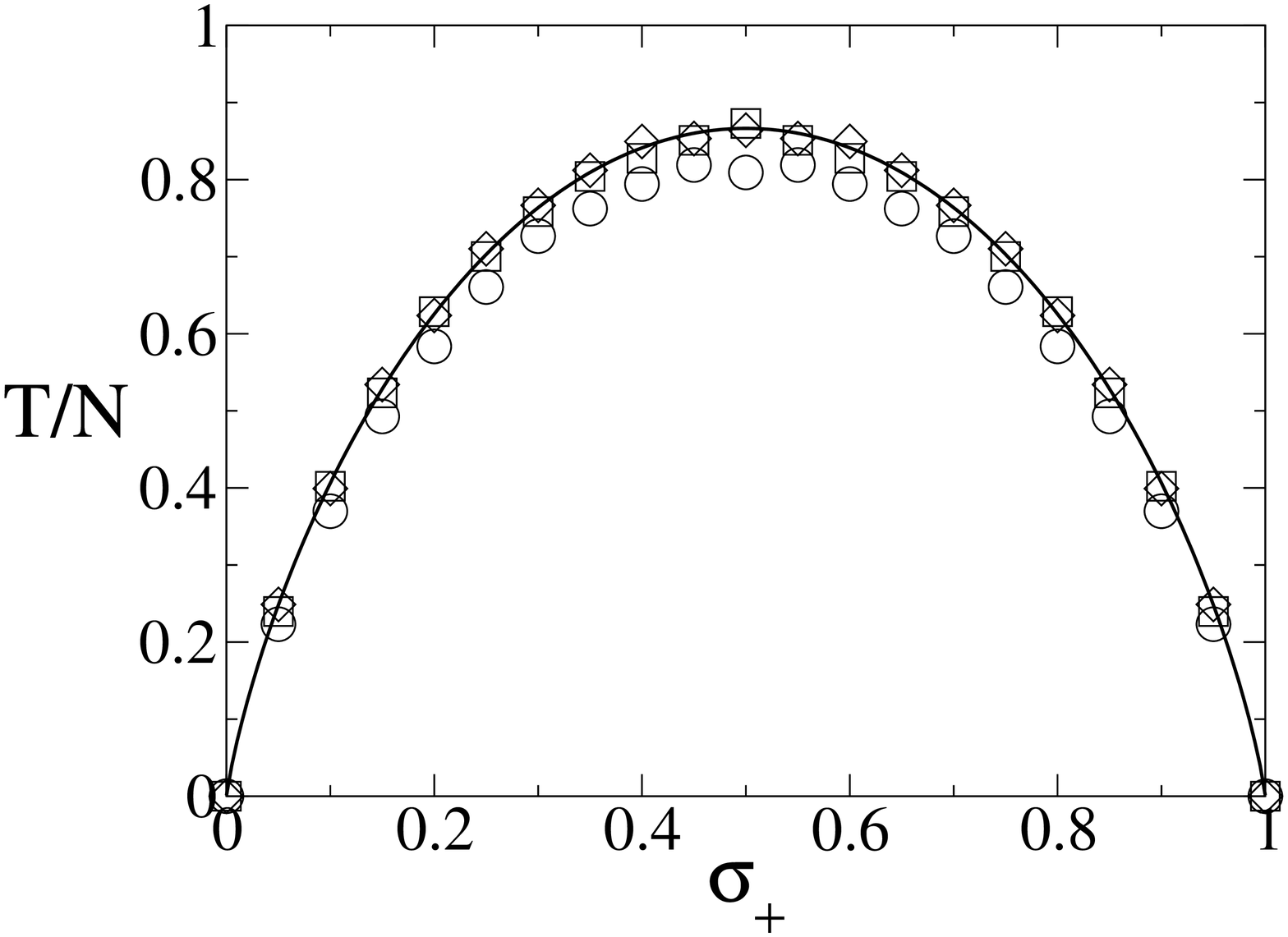} &
\includegraphics[width=3.0in]{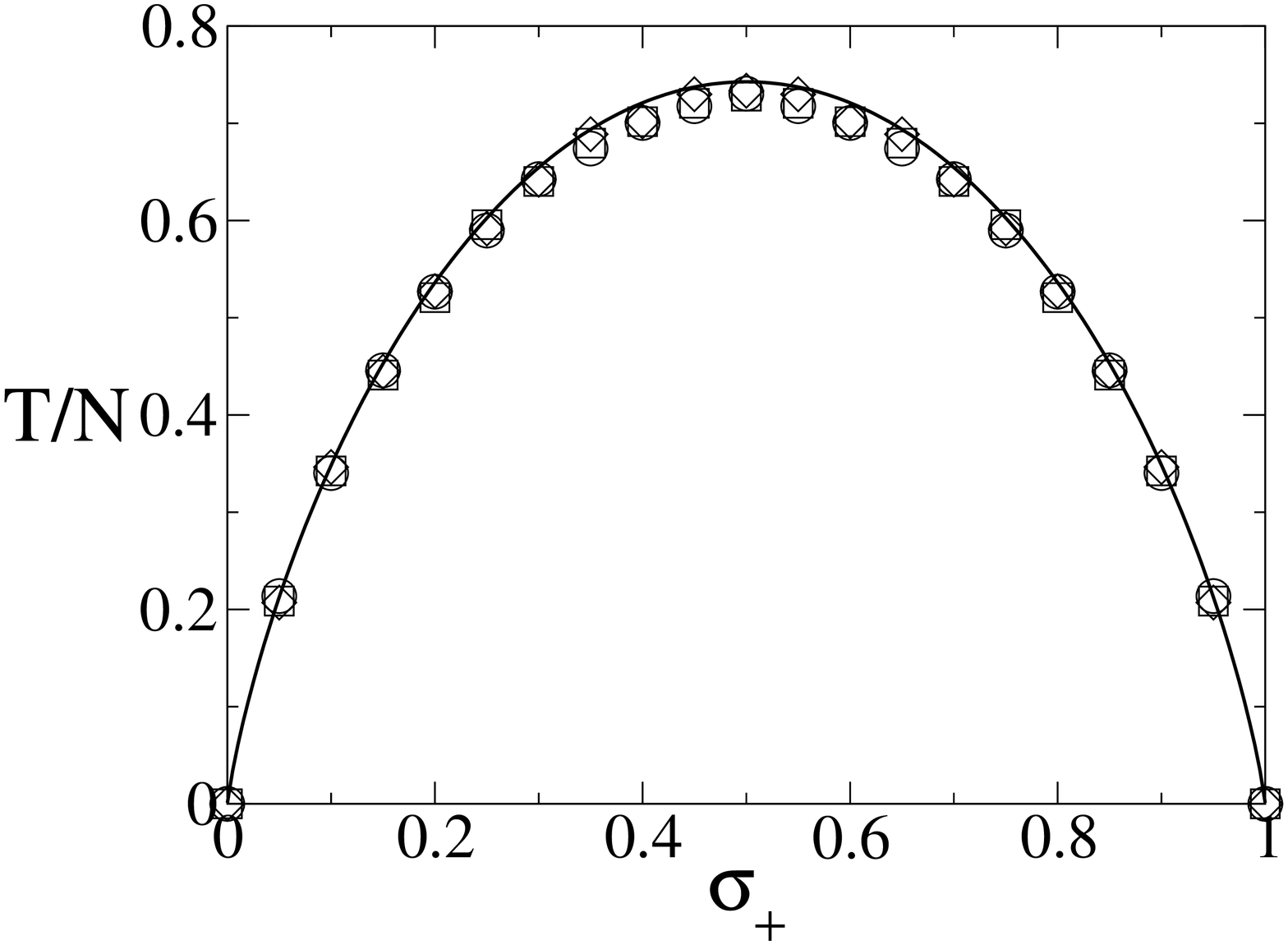} \\ 
\includegraphics[width=3.0in]{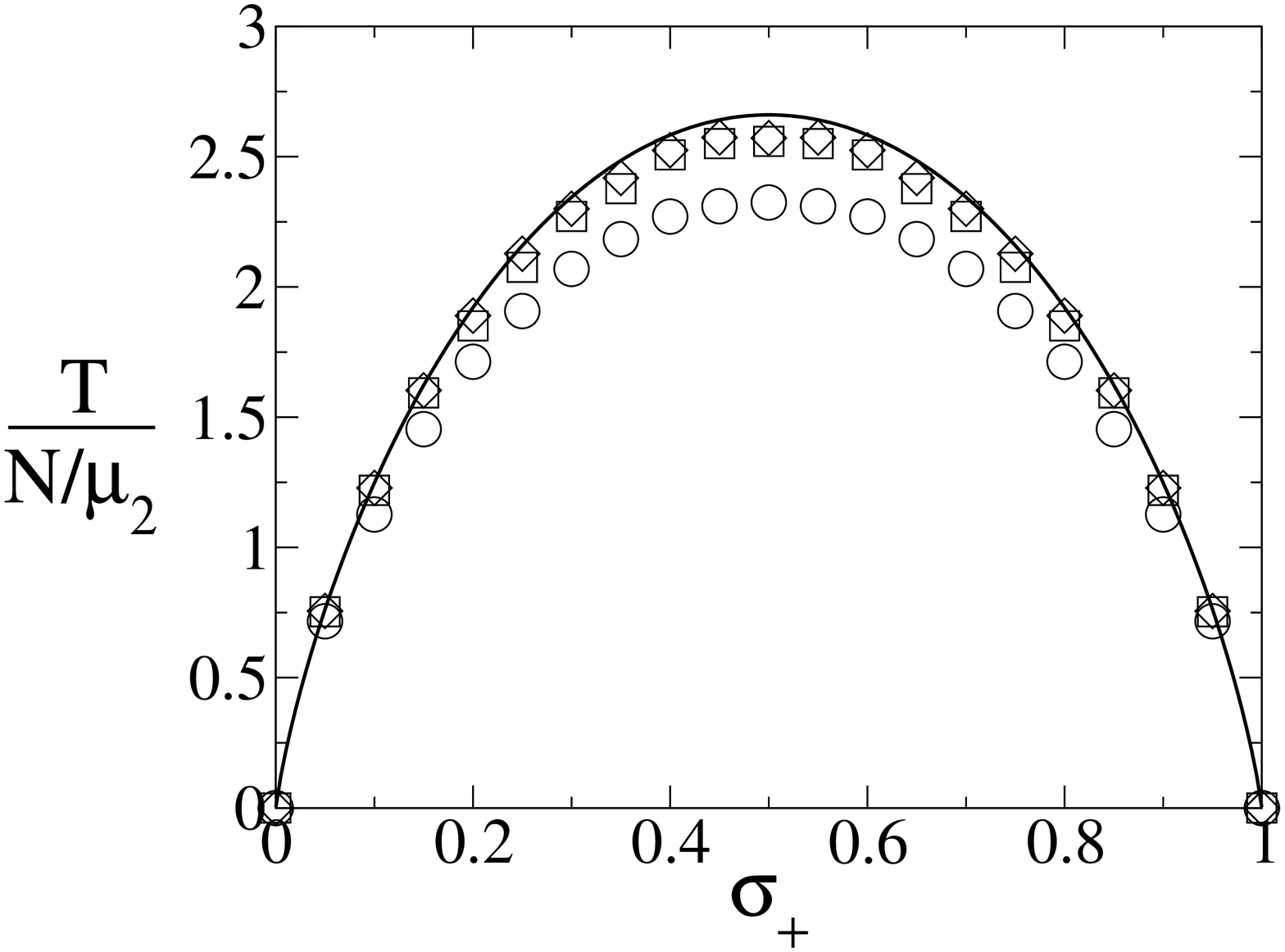} &
\end{array}$
\end{center}
\caption{Scaled ordering times vs initial density of $+$ spins $\sigma_+$ for 
  networks of size $N=10^2$ (circles), $N=10^3$ (squares) and $N=10^4$ 
  (diamonds).  Plots correspond to DR (top-left) and ER networks
  (top-right) with average degree $\mu=4$ and BA networks (bottom-left) with
  $\mu = 20$.  Solid lines are the analytical predictions from 
  Eq.~(\ref{tau-sigma}).}
\label{Tau}
\end{figure}

This expression differs with the one obtained in work \cite{Sood05} by a 
prefactor of $\frac{\mu-1}{\mu-2}$.  However this factor does not seem to
change the scaling of $T(m)$ with the system size $N$, that was found to be in
good agreement with numerical simulations.  In Fig.~(\ref{Tau}) we show the 
ordering time $t(\sigma_+)$ as a function of the initial density of $+$ spins, 
for a BA network with $\mu=20$, ER and DR networks with $\mu=6$.

For a fixed $N$, Eq.~(\ref{tau-sigma}) predicts that $T(m)$ diverges at
$\mu=2$, but ordering times in the voter model are finite for finite sizes.
To analyze this point, we numerically calculated $T$ for an Erd\H{o}s-R\'enyi
network as function of $\mu$ for initial densities $\sigma_+=\sigma_-=1/2$
(see Fig.~\ref{Tau-k}).  For low values of $\mu$, there is a fraction of nodes
with zero degree that have no  dynamics, thus we normalized $T$ by the number
of nodes $N$ with degree larger  than zero.  As we observe in
Fig.~\ref{Tau-k},  when $\mu$ decreases the  analytical solution given by
Eq.~(\ref{tau-sigma}) with $\mu_2=\mu(\mu+1)$  start to diverge from the
numerical solution.  This disagreement might be due to the fact that our
mean-field approach assumes that the system is homogeneous, and neglects every
sort of  fluctuations, which are important in networks with low connectivity.
However, we still find that $T$ reaches a maximum at $\mu \simeq 2$, where it 
seems to grow faster than $N$.

\begin{figure}[t]
\begin{center}
 \includegraphics[width=0.8\textwidth]{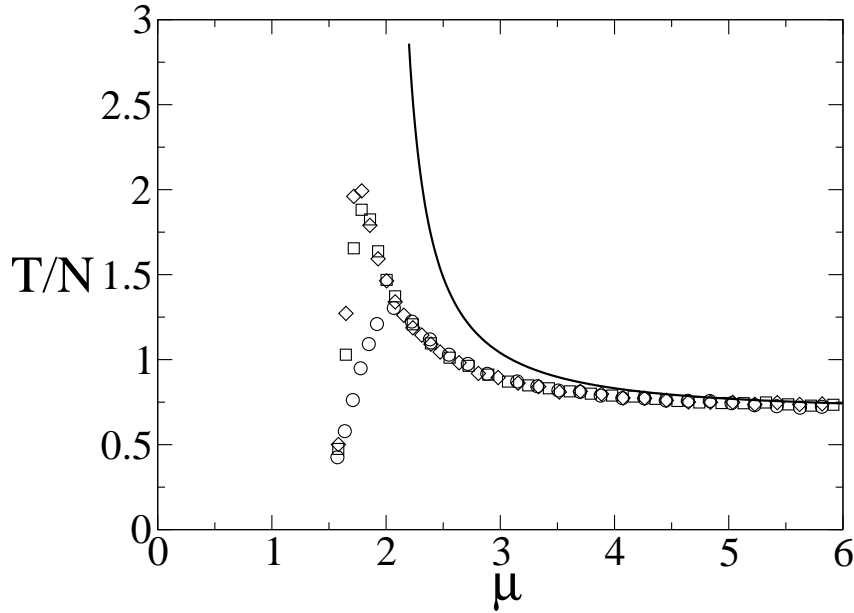}
\end{center}
\caption{Scaled ordering times vs average degree $\mu$ for Erd\H{o}s-R\'enyi
 networks  with $N=100$ (circles), $N=1000$ (squares) and $N=2000$ (diamonds)
 nodes.  The system size $N$ was taken as the number of nodes in the network
 with  degree larger than zero.   The initial spin densities were
 $\sigma_+=\sigma_-=1/2$.  The solid line is the solution given by
 Eq.~(\ref{tau-sigma}).}
\label{Tau-k}
\end{figure}

\section{Summary and conclusions}
\label{summary}

In this article we have presented a mean-field approach over the density of
active links that provides a description of the time evolution and final
states of the voter model on heterogenous networks in both infinite and finite
systems.  The theory gives analytical results that are in good agreement with
simulations of the model and also shows the connection between previous
numerical and analytical results.  The relation between the density of active
links $\rho$ and the density of $+$ spins $\sigma_+$ expressed in
Eq.~(\ref{r-t1}) allows to treat random graphs as complete graphs, and to find
expressions for $\rho$ and the mean ordering time in finite systems.  For
large average degree values, Eq.~(\ref{r-t1}) reduces to the expression  for
the density of active links in complete graph.  Therefore, this work confirms
that disordered networks with large enough connectivity are mean-field in
character for the dynamics of the voter model.

We find that when the average degree $\mu$ is smaller than $2$, the system
orders, while for $\mu>2$, the average density of active links in surviving
runs reaches a plateau of height  $\frac{(\mu-2)}{3(\mu-1)}$.  Due to
fluctuations, a finite system always falls into an absorbing, fully-ordered
state.  The relaxation time $T$ to  the final absorbing state scales with
the system size $N$ and the first and  second moments, $\mu$ and $\mu_2$
respectively, of the degree  distribution, as \\ $T \sim \frac{(\mu-1)
\mu^2 N}{(\mu-2)\,\mu_2}$.

Plateaus are also found on correlated networks with some level of node degree
correlations, like for instance on small-world
\cite{Castellano03,Suchecki05a}, even though the  plateau is lower  than the
one predicted by our theory.  It might be interesting to modified  the
mean-field approach to account for degree correlations that correctly
reproduce the behavior in very general networks.

We would like to acknowledge financial support from MEC (Spain), CSIC (Spain)
and EU through  projects FISICOS, PIE200750I016 and PATRES respectively.

\clearpage

\appendix

\section{Average density of active links}
\label{ave-den}

To integrate Eq.(\ref{integral}), we use the series expansion Eq.(\ref{P-t}) 
for $P(m,t')$ and write
\begin{equation}
\label{rho-ave}
\langle \rho(t') \rangle = 
\xi \sum_{l=0}^\infty A_l \,D_l \,e^{-(l+1) (l+2)\,t'},
\end{equation}   
where we define the coefficient
\begin{eqnarray*}
D_l \equiv \int_{-1}^1 dm\, (1-m^2)\, C_l^{3/2}(m).
\end{eqnarray*}
To obtain the coefficients $A_l$, we assume that the initial magnetization is
$m(t=0) = m_0$, i.e., $P(m,t=0)=\delta(m-m_0)$, from where the expansion for
$P(m,t')$ becomes
\begin{eqnarray*}
\sum_{l=0}^{\infty} A_l \, C_l^{3/2}(m) = \delta(m-m_0).
\end{eqnarray*}
Multiplying both sides of the above equation by $(1-m^2)\,C_{l'}^{3/2}(m)$ and 
integrating over $m$ gives
\begin{equation}
\label{Al}
\sum_{l=0}^{\infty} \frac{2 (l+1)(l+2)}{(2 l+3)}\,A_l\, \delta_{l,l'} =
(1-m_0^2)\,C_{l'}^{3/2}(m_0)
\end{equation}
where we used the orthogonality relation for the Gegenbauer polynomials 
Eq.~MS 5.3.2 (8) in page 983 of \cite{Grandshteyn} with $\lambda=3/2$
\begin{equation}
\label{ortho}
\int_{-1}^1 dm \,C_l^{3/2}(m)\,C_{l'}^{3/2}(m)\,(1-m^2) = 
\frac{\pi \,\Gamma(l+3)}{4\, l!\, (l+3/2) [\Gamma(3/2)]^2} \delta_{l,l'} 
\end{equation}
and the identities $\Gamma(l)=(l-1)!$, $\Gamma(l+1)=l\, \Gamma(l)$ 
and $\Gamma(1/2)= \sqrt{\pi}$.
Then, from Eq.~(\ref{Al}) we obtain
\begin{equation}
\label{Al2}
A_l = \frac{(2l+3)(1-m_0^2)\,C_l^{3/2}(m_0)}{2(l+1)(l+2)}.
\end{equation}

To find $D_l$, we use that the zeroth order polynomial is $C_0^{3/2}(m)=1$,
together with the orthogonality relation Eq.~(\ref{ortho}):
\begin{eqnarray}
\label{Dl}
D_l &=& \int_{-1}^{1} dm\, C_l^{3/2}(m)\,C_0^{3/2}(m)\,(1-m^2) = 
\frac{\pi \Gamma(l+3)}{4\, l!\, (l+3/2) [\Gamma(3/2)]^2} \delta_{l,0}
\nonumber \\
&=&\frac{2(l+1)(l+2)}{(2 l+3)} \delta_{l,0}.
\end{eqnarray}
Then, using Eqns.~(\ref{Al2}) and (\ref{Dl}) we find that the coefficients
$A_l$ and $D_l$ are related by 
$A_l\,D_l=(1-m_0^2)\,C_l^{3/2}(m_0)\,\delta_{l,0}$.  Replacing this relation in
Eq.~(\ref{rho-ave}) and performing the summation we finally obtain
\begin{eqnarray*}
\langle \rho(t') \rangle = \xi \,(1-m_0^2)\,e^{-2\,t'},
\end{eqnarray*}
as quoted in Eq.~(\ref{rho-ave1}).

\section{Calculation of $\mu_2$ for Bar\'abasi-Albert networks}
\label{mu2-tau}

The Bar\'abasi-Albert network is generated by starting with a number $m$ of
nodes, and adding, at each time step, a new node with $m$ links that connect
to $m$  different nodes in the network.  When the number of nodes in the
system is  $N$, the total number of links is $m N$, and therefore the average
degree is  $\mu = 2 m$.  The expression for the resulting degree distribution,
calculated  for instance in \cite{Dorogovstev00}, as a function of $\mu$ is
\begin{equation} 
P(k)=\frac{\mu(\mu+2)}{2 k (k+1) (k+2)},        
\end{equation}
and its second moment is
\begin{eqnarray}
\label{mu2}
\mu_2 &=& \int_{\mu/2}^{k_{max}} k^2 P(k) dk=
\frac{\mu(\mu+2)}{2} \int_{\mu/2}^{k_{max}} \frac{k \,dk}{(k+1)(k+2)} \\
\nonumber          
&=&\frac{\mu(\mu+2)}{2} 
\ln \left[ \frac{2(k_{max}+2)^2 (\mu+2)}{(k_{max}+1) (\mu+4)^2} \right].
\end{eqnarray}   
The lower limit $\mu/2$ of the above integrals correspond to the lowest
possible degree $m$, since nodes already have $m$ links when they are added to
the network.  The reason for an upper limit $k_{max}$ is that the contribution 
to $\mu_2$ from large degree terms is important due to the slow asymptotic 
decay $P(k) \sim k^{-3}$, unlike for instance in Erd\H{o}s-R\'enyi or 
Exponential
networks where $P(k)$ decays faster than $k^{-3}$, thus high degree terms 
become irrelevant.  $k_{max}$ is estimated as the degree for which the number 
of nodes with degree larger than $k_{max}$ is less than one.  Then
\begin{eqnarray*}
\frac{1}{N} = 
\frac{\mu(\mu+2)}{2} \int_{k_{max}}^{\infty} \frac{dk}{k(k+1) (k+2)}=
\frac{\mu(\mu+2)}{4} 
\ln \left( \frac{(k_{max}+1)^2}{k_{max}(k_{max}+2)} \right).        
\end{eqnarray*}
Assuming $k_{max} \gg 1$, the expansion of the logarithm to first order in 
$1/k_{max}$ is $1/k_{max}^2$. Then, solving for $k_{max}$, we obtain 
\begin{equation}
\label{kmax}
k_{max} \simeq \sqrt{u(u+2)/4} N^{1/2},
\end{equation}
i.e, the maximum degree diverges with the system size.

Taking $k_{max} \gg 1$ in Eq.~(\ref{mu2}) and replacing the value 
of $k_{max}$ from Eq.~(\ref{kmax}) gives the expression quoted in 
table \ref{table} for the second moment of a BA network
\begin{equation} 
\mu_2 = \frac{\mu (\mu+2)}{4}\ln\left(\frac{\mu(\mu+2)^3 N}{(\mu+4)^4}\right).
\end{equation}

\section{Survival probability}
\label{surv}

By using the series representation Eq.~(\ref{P-t}), the survival 
probability quoted in Eq.(\ref{S-t2}) can be written as 
\begin{equation}
\label{S-t3}
S(t) = \sum_{l=0}^{\infty} A_l\,B_l \,e^{-(l+1) (l+2)\,t'},
\end{equation}
where we define 
\begin{equation}
B_l \equiv \int_{-1}^{1} dm \,C_l^{3/2}(m).
\end{equation}
To obtain the coefficients $B_l$, we use the derivative identity 
$C_l^{3/2}(m) = \frac{d}{dm}C_{l+1}^{1/2}(m)$ derived from Eq.~MS 5.3.2~(1) 
in page 983 of \cite{Grandshteyn} with $\lambda=3/2$.  Then
\begin{eqnarray}
\label{Bl}
  B_l = C_{l+1}^{1/2}(1)-C_{l+1}^{1/2}(-1) = 1 - (-1)^{l+1} 
= \left\{ \begin{array}{ll}
         0 & \mbox{$l$ odd} \\
         2 & \mbox{$l$ even} \end{array} \right.
\end{eqnarray}
where we have used the relations $C_l^{1/2}(1)=1$ $\forall\, l$ and 
$C_l^{1/2}(-1)=(-1)^l$ that follow from Eq.~MO 98 (4) (page 983) and the 
parity of the polynomials (page 980) of \cite{Grandshteyn} respectively.

An explicit function for the coefficients $A_l$ of Eq.~(\ref{Al2}) can only be 
found for the $m_0=0$ case, given that for $m_0 \not= 0$ it seems that a 
closed expression for the polynomials $C_l^{3/2}(m_0)$ cannot be obtained. 
To obtain the coefficients $C_l^{3/2}(0)$ we use the recursion relation 
Eq.~Mo 98 (4) (page 981) of \cite{Grandshteyn} for $m \equiv x=0$ and 
$\lambda=3/2$, together with the values of the zeroth and first order
polynomials $C_0^{3/2}(0)=1$ and $C_1^{3/2}(0)=0$.  Then
\begin{equation}
C_l^{3/2}(0) = -\frac{(l+1)}{l}\,C_{l-2}^{3/2}(0)=
\left\{ \begin{array}{ll}
         0 & \mbox{$l$ odd} \\
         (-1)^{l/2}\, \frac{(l+1)!!}{l\,!!}&\mbox{$l$ even} \end{array}\right.
\end{equation} 
Plugging the above expression into Eq.(\ref{Al2}) gives $A_l=0$ for 
$l$ odd and \\$A_l = \frac{(-1)^{l/2} (2l+3)(l-1)!!}{2(l+2)!!}$ for $l$ even.

Then, using Eq.~(\ref{Bl}), the product $A_l\,B_l$ can be written as 
\begin{equation}
A_l\,B_l = 
\left\{ \begin{array}{ll}
         0 & \mbox{$l$ odd} \\
     \frac{(-1)^{l/2}(2l+3)(l-1)!!}{(l+2)!!}&\mbox{$l$ even}\end{array}\right.
\end{equation} 
Finally, making the variable change $l \to 2\, l$, Eq.~(\ref{S-t3}) becomes
\begin{equation}
\label{S-t4}
S(t') = \sum_{l=0}^{\infty} \frac{(-1)^{l}(4l+3)(2l-1)!!}{(2l+2)!!} 
\,e^{-2(2l+1) (l+1)\,t'}.
\end{equation}
Replacing $t'$ by $t/\tau(\mu,N)$, we obtain the expression quoted in
Eq.~(\ref{S-t0}).

\end{document}